\documentclass[runningheads]{llncs} 
\usepackage{algo}
\usepackage[hidelinks]{hyperref}
\hypersetup{colorlinks=true,linkcolor=blue,citecolor=blue,urlcolor=blue}

\title{On the Formalization of \\ the Notion of an Algorithm}
\author{C.A. Middelburg \\
        {\small ORCID: \url{https://orcid.org/0000-0002-8725-0197}}}
\institute{Informatics Institute, Faculty of Science, University of
           Amsterdam \\
           Science Park~900, 1098~XH Amsterdam, the Netherlands \\
           \email{C.A.Middelburg@uva.nl}}

\titlerunning{On the Formalization of the Notion of an Algorithm}
\authorrunning{C.A. Middelburg}

\begin{document}

\maketitle

\begin{abstract}
The starting point of this paper is a collection of properties of an
algorithm that have been distilled from the informal descriptions of 
what an algorithm is that are given in standard works from the 
mathematical and computer science literature.
Based on that, the notion of a proto-algorithm is introduced. 
The thought is that algorithms are equivalence classes of 
proto-algorithms under some equivalence relation.
Three equivalence relations are defined.
Two of them give bounds between which an appropriate equivalence 
relation must lie.
The third lies in between these two and is likely an appropriate 
equivalence relation.
A sound method is presented to prove, using an imperative process 
algebra based on ACP, that this equivalence relation holds between two 
proto-algorithms.

\end{abstract}

\section{Introduction}
\label{sect-intro}

In many works from the mathematical and computer science literature, 
including standard works such as~\cite{Kle67a,Knu97a,Mal70a,Rog67a}, the 
notion of an algorithm is informally characterized by properties that 
are considered the most important ones of an algorithm.
Most of those characterizations agree with each other and indicate that
an algorithm is considered to express a pattern of behaviour by which 
all instances of a computational problem can be solved.
A remark like ``Formally, an algorithm is a Turing machine'' is often 
made in the works concerned if additionally Turing machines are 
rigorously defined.
 
However, the viewpoint that the formal notion of a Turing machine is a
formalization of the intuitive notion of an algorithm is unsatisfactory
in at least two ways:
(a)~a Turing machine expresses primarily a way in which a computational 
problem-solving pattern of behaviour can be generated and
(b)~a Turing machine restricts the data involved in such a pattern of 
behaviour to strings over some finite set of symbols.
There are not many alternative formalizations of the notion of an 
algorithm that are regularly cited.
To the best of my knowledge, the main exceptions are the ones that can 
be found in~\cite{Gur00a,MP08a}. 
In both papers, a notion of an algorithm is formally defined that does 
not depend on a particular machine model such as the Turing machine 
model. 

In~\cite{MP08a}, an algorithm is defined as a fairly complex 
set-theoretic object.
The definition has its origins in the idea that, if a partial function 
is defined recursively by a system of equations, that system of 
equations induces an algorithm.
An algorithm according to this definition fails to have many properties 
that are generally considered to belong to the most important ones of an 
algorithm.

In~\cite{Gur00a}, an algorithm is defined as an object that satisfies 
certain postulates.
The postulates concerned appear to be devised with the purpose that 
Gurevich’s abstract state machines would satisfy them. 
However, this definition covers objects that have almost all properties 
that are generally considered to belong to the most important ones of an 
algorithm as well as more abstract objects that have almost none of 
those properties.

What is mentioned above about the formalizations of the notion of an
algorithm in~\cite{Gur00a,MP08a} makes them unsatisfactory as well.
This state of affairs motivated me to start a quest for a formalization 
of the notion of an algorithm that is more satisfactory than the 
existing ones.
One possibility is to investigate whether this can be done by adapting 
the postulates from~\cite{Gur00a} or adding postulates to them.
Another possibility is to investigate whether a constructive definition 
can be given.
This is what will be done in this paper.
In addition, the connection between the resulting objects and the 
processes considered in the imperative process algebra presented 
in~\cite{Mid21a} will be investigated.

In~\cite{BM14a}, I made a first attempt to give a constructive 
definition.
A main drawback of the approach followed there is that the data 
involved in an algorithm is restricted to bit strings.
The idea was that this restriction could be discarded without much 
effort.
This turned out not to be the case.
Therefore, I follow a rather different approach in this paper.

\section{The Informal Notion of an Algorithm}
\label{sect-algo-informal}

What is an algorithm?
A brief answer to this question usually goes something like this: an 
algorithm is a procedure for solving a computational problem in a finite 
number of steps.
This is a reasonable answer. 
A difficulty is that it is common to describe a computational problem 
informally as a problem that can be solved using an algorithm. 
For this reason, first a description of a computational problem that 
does not refer to the notion of an algorithm must be given:
\begin{quote}
A computational problem is a problem where, given an input value that 
belongs to a certain set, an output value that is in a certain relation 
to the given input value must be found if it exists.
The input values that belong to the certain set are also called the 
instances of the problem and an output value that is in the certain 
relation to the given input value is also called a solution for the 
instance concerned.
\end{quote}

The existing viewpoints on what an algorithm is indicate that something 
like the following properties are essential for an algorithm: 
\begin{itemize}
\item
an algorithm is a finite expression of a pattern of behaviour by which 
all instances of a computational problem can be solved;
\item
the pattern of behaviour expressed by an algorithm is made up of 
discrete steps, each of which consists of performing an elementary 
operation or inspecting an elementary condition unless it is the initial 
step or a final step;
\item
the pattern of behaviour expressed by an algorithm is such that there is 
one possible step immediately following a step that consists of 
performing an operation;
\item
the pattern of behaviour expressed by an algorithm is such that there is 
one possible step immediately following a step that consists of 
inspecting a condition for each outcome of the inspection;
\item
the pattern of behaviour expressed by an algorithm is such that the 
initial step consists of inputting an input value of the problem 
concerned;
\item
the pattern of behaviour expressed by an algorithm is such that, for 
each input value of the problem concerned for which a correct output
value exists, a final step is reached after a finite number of steps and 
that final step consists of outputting a correct output value for that 
input value;
\item
the steps involved in the pattern of behaviour expressed by an algorithm
are precisely and unambiguously defined and can be performed exactly in 
a finite amount of time.
\end{itemize}
These properties give an intuitive characterization of the notion of an 
algorithm and form the starting point for the formalization of this 
notion in upcoming sections. 
They have been distilled from the descriptions of what an algorithm is
that are given in standard works from the mathematical and computer 
science literature such as~\cite{Kle67a,Knu97a,Mal70a,Rog67a}.
They can also be found elsewhere in the mathematical and computer 
science literature and even in the philosophical literature on 
algorithms, see e.g.~\cite{Hil16a,Pap23a}.

Usually it is also mentioned in some detail how an algorithm is 
generally expressed.
However, usually it is mentioned at most in passing that an 
algorithm expresses a pattern of behaviour.
Following~\cite{Dij71a}, this point is central here.
The reason for this is that, in order to formalize the notion of an 
algorithm well, it is more important to know what an algorithm expresses 
than how an algorithm is expressed.

Recently, discussions about the notion of an algorithm take also place 
in the social sciences.
This leads to viewpoints on algorithms that are useless in mathematics 
and computer science. 
For example, in~\cite{Sea17a} is proposed to view algorithms as 
`heterogeneous and diffuse sociotechnical systems'.
Such viewpoints preclude formalization and are therefore disregarded.

It should be noted that the characterization of the notion of an 
algorithm given by the above-mentioned properties of an algorithm 
reflects a rather operational view of what an algorithm is.
In a more abstract view of what an algorithm is, an algorithm expresses 
a collection of patterns of behaviour that are equivalent in some 
well-defined way.
We will come back to this at the end of Section~\ref{sect-proto-algo}.

\section{Proto-Algorithms}
\label{sect-proto-algo}

In this section, the notion of an proto-algorithm is introduced.
The thought is that algorithms are equivalence classes of 
proto-algorithms under an appropriate equivalence relation.
An equivalence relation that is likely an appropriate one is introduced
in Section~\ref{sect-algorithm}.

The notion of a proto-algorithm will be defined in terms of three 
auxiliary notions. 
The definition of one of these auxiliary notions is based on the 
well-known notion of a rooted labeled directed graph.
However, the definitions of this notion given in the mathematical and 
computer science literature vary. 
Therefore, the definition that is used in this paper is given first.

\begin{udef}
A \emph{rooted labeled directed graph} $G$ is a sextuple 
$(V,E,\LBLv,\LBLe,l,r)$, where:
\begin{itemize}
\item
$V$ is a non-empty finite set, whose members are called the 
\emph{vertices} of $G$;
\item
$E$ is a subset of $V \x V$, whose members are called the 
\emph{edges} of $G$;
\item
$\LBLv$ is a countable set, whose members are called the 
\emph{vertex labels} of $G$; 
\item
$\LBLe$ is a countable set, whose members are called the 
\emph{edge labels} of $G$; 
\item
$l$ is a partial function from $V \union E$ to $\LBLv \union \LBLe$ 
such that
\begin{itemize}
\item[]
for all $v \in V$ for which $l(v)$ is defined, $l(v) \in \LBLv$ and
\item[]
for all $e \in E$ for which $l(e)$\, is defined, $l(e) \in \LBLe$,
\end{itemize}
called the \emph{labeling function} of $G$;
\item 
$r \in V$, called the \emph{root} of $G$.
\end{itemize}
\end{udef}
The additional graph theoretical notions defined below are also used in 
this paper.
\begin{udef}
Let $G = (V,E,\LBLv,\LBLe,l,r)$ be a rooted labeled directed graph.
Then a \emph{cycle} in $G$ is a sequence 
$v_1\, \ldots\, v_{n+1} \in V^*$
such that, for all $i \in \set{1,\ldots,n}$, $(v_i,v_{i+1}) \in E$,
$\mathrm{card}(\set{v_1,\ldots,v_n}) = n$, and $v_1 = v_{n+1}$.
Let, moreover, $v \in V$.
Then 
the \emph{indegree} of $v$, written $\indeg(v)$, is  
$\mathrm{card}(\set{v' \where (v',v) \in E})$ 
and
the \emph{outdegree} of $v$, written $\outdeg(v)$, is  
$\mathrm{card}(\set{v' \where (v,v') \in E})$. 
\end{udef}

We proceed with defining the three auxiliary notions, starting with the 
notion of an alphabet.
This notion concerns the symbols used to refer to the operations and 
conditions involved in the steps of which the pattern of behaviour 
expressed by an algorithm is made up.

\begin{udef}
An \emph{alphabet} $\Sigma$ is a couple $(F,P)$, where:
\begin{itemize}
\item
$F$ is a countable set, whose members are called the 
\emph{function symbols} of $\Sigma$;
\item
$P$ is a countable set, whose members are called the 
\emph{predicate symbols} of $\Sigma$;
\item
$F$ and $P$ are disjoint sets and $\ini,\fin \in F$.
\end{itemize}
\end{udef}
We write $\widetilde{F}$, where $F$ is the set of function symbols of an 
alphabet, for the set $F \diff \set{\ini,\fin}$.

The function symbols and predicate symbols of an alphabet refer to the 
operations and conditions, respectively, involved in the steps of which 
the pattern of behaviour expressed by an algorithm is made up.
The function symbols $\ini$ and $\fin$ refer to inputting an input value 
and outputting an output value, respectively.

We are now ready to define the notions of a $\Sigma$-algorithm graph and
a $\Sigma$-interpretation.
They concern the pattern of behaviour expressed by an 
algorithm.

\begin{udef}
Let $\Sigma = (F,P)$ be an alphabet.
Then a \emph{$\Sigma$-algorithm graph} $G$ is a rooted labeled directed 
graph $(V,E,\LBLv,\LBLe,l,r)$ such that
\begin{itemize}
\item
$\LBLv = F \union P$;
\item
$\LBLe = \set{0,1}$;
\item
for all $v \in V$:
\begin{itemize}
\item
$l(v) = \ini$ iff $v = r$;
\item
if $l(v) = \ini$, then
$\indeg(v) = 0$, $\outdeg(v) = 1$, and, for the unique $v' \in V$ for 
which $(v,v') \in E$, $l((v,v'))$ is undefined;
\item
if $l(v) = \fin$, then
$\indeg(v) > 0$ and $\outdeg(v) = 0$;
\item
if $l(v)  \in \widetilde{F}$, then
$\indeg(v) > 0$, $\outdeg(v) = 1$, and, for the unique $v' \in V$ for 
which $(v,v') \in E$, $l((v,v'))$ is undefined;
\item
if $l(v)  \in P$, then
$\indeg(v) > 0$, $\outdeg(v) = 2$, and, for the unique $v' \in V$ and 
$v'' \in V$ with $v' \neq v''$ for which $(v,v') \in E$ and 
$(v,v'') \in E$, $l((v,v'))$ is defined, $l((v,v''))$ is defined, and
$l((v,v')) \neq l((v,v''))$;
\end{itemize}
\item
if $v_1\, \ldots\, v_{n+1}$ is a cycle in $G$, then,
for some $v \in \set{v_1,\ldots,v_n}$, $l(v) \in F$.
\end{itemize}
\end{udef}
$\Sigma$-algorithm graphs are somewhat reminiscent of program schemes 
as defined, for example, in~\cite{Wey79a}.

In the above definition, the condition on cycles in a $\Sigma$ algorithm 
graph excludes infinitely many consecutive steps, each of which consists 
of inspecting a condition.

In the above definition, the conditions regarding the vertices of a 
$\Sigma$-algorithm graph correspond to the essential properties of an 
algorithm mentioned in Section~\ref{sect-algo-informal} that concern its 
structure.
Adding an interpretation of the symbols of the alphabet $\Sigma$ to a
$\Sigma$-algorithm graph yields something that has all of the mentioned 
essential properties of an algorithm.

\begin{udef}
Let $\Sigma = (F,P)$ be an alphabet.
Then a \emph{$\Sigma$-interpretation} $\cI$ is a quadruple 
$(D,\Din,\Dout,I)$, 
where:
\begin{itemize}
\item
$D$ is a set, called the \emph{main domain} of $\cI$;
\item
$\Din$ is a set, called the \emph{input domain} of $\cI$;
\item
$\Dout$ is a set, called the \emph{output domain} of $\cI$;
\item
$I$ is a total function from $F \union P$ to the set of all total 
computable functions from $\Din$ to $D$, $D$ to $\Dout$, $D$ to $D$ or 
$D$ to $\set{0,1}$ such that:
\begin{itemize}
\item
$I(\ini)$ is a function from $\Din$ to $D$;
\item
$I(\fin)$ is a function from $D$ to $\Dout$;
\item
for all $f \in \widetilde{F}$, $I(f)$ is a function from $D$ 
to $D$;
\item
for all $p \in P$, $I(p)$ is a function from $D$ to 
$\set{0,1}$;
\end{itemize}
\item
there does not exist a $D' \subset D$ such that:
\begin{itemize}
\item
for all $d \in \Din$, $I(\ini)(d) \in D'$;
\item
for all $f \in \widetilde{F}$, for all $d \in D'$, $I(f)(d) \in D'$.
\pagebreak[2]
\end{itemize}
\end{itemize}
\end{udef}
In the above definition, the minimality condition on $D$ is not 
essential, but this condition facilitates establishing a connection 
between proto-algorithms and the processes considered in the imperative 
process algebra \deBPAde\ (see Section~\ref{sect-algo-procs}).

The pattern of behavior expressed by an algorithm can completely be 
represented by the combination of an alphabet $\Sigma$, a 
$\Sigma$-algorithm graph $G$, and a $\Sigma$-interpretation $\cI$.
This brings us to defining the notion of a proto-algorithm. 

\begin{udef}
A \emph{proto-algorithm} $A$ is a triple $(\Sigma,G,\cI)$, where:
\begin{itemize}
\item
$\Sigma$ is an alphabet, called the \emph{alphabet} of $A$;
\item
$G$ is a $\Sigma$-algorithm graph, called the \emph{algorithm graph} of 
$A$;
\item
$\cI$ is a $\Sigma$-interpretation, called the \emph{interpretation} of 
$A$.
\end{itemize}
\end{udef}

\sloppy
Let $A = (\Sigma,G,\cI)$ be a proto-algorithm, where $\Sigma = (F,P)$, 
$G = (V,E,\LBLv,\LBLe,\linebreak[2]l,r)$, and $\cI = (D,\Din,\Dout,I)$.
Then the intuition is that $A$ is something that goes through states, 
where states are elements of the set 
$\Din \union (V \x D) \union \Dout$.
The elements of $\Din$, $V \x D$, and $\Dout$ are called input states, 
internal states, and output states, respectively.
$A$ goes from one state to the next state by making a step, it starts in 
an input state, and it stops in an output state.
The state that $A$ is in determines what the step to the next state
consists of and what the next state is as follows: 
\begin{itemize}
\item
if $A$ is in input state $d$, then the step to the next state consists 
of applying function $I(\ini)$ to $d$ and the next state is the unique 
internal state $(v',d')$ such that $(r,v') \in E$, and 
$I(\ini)(d) = d'$;
\item
if $A$ is in internal state $(v,d)$ and 
$l(v) \in \widetilde{F}$, then the step to the next state 
consists of applying function $I(l(v))$ to $d$ and the next state is 
the unique internal state $(v',d')$ such that $(v,v') \in E$, and 
$I(l(v))(d) = d'$;
\item
if $A$ is in internal state $(v,d)$ and $l(v) \in P$, then the step to 
the next state consists of applying function $I(l(v))$ to $d$ and the 
next state is the unique internal state $(v',d)$ such that 
$(v,v') \in E$, and $I(l(v))(d) = l((v,v'))$;
\item
if $A$ is in internal state $(v,d)$ and $l(v) = \fin$, then the step to 
the next state consists of applying function $I(\fin)$ to $d$ and the 
next state is the unique output state $d'$ such that $I(\fin)(d) = d'$.
\end{itemize}
This informal explanation of how the state that $A$ is in determines 
what the next state is, is formalized by the algorithmic step function
$\astep_A$ defined in Section~\ref{sect-algorithm}. 

The term proto-algorithm has been chosen instead of the term algorithm  
because proto-algorithms are considered too concrete to be called 
algorithms.
For example, from a mathematical point of view, it is natural to 
consider the behavioral patterns expressed by isomorphic 
proto-algorithms to be the same.
Isomorphism of proto-algorithms is defined as expected.

\begin{udef}
Let $A = (\Sigma,G,\cI)$ and $A' = (\Sigma',G',\cI')$ be 
proto-algorithms, where
$\Sigma = (F,P)$, $\Sigma' = (F',P')$, 
$G = (V,E,\LBLv,\LBLe,l,r)$, $G' = (V',E',\LBLv',\LBLe',l',r')$, 
$\cI = (D,\Din,\Dout,I)$, and $\cI' = (D',\Din',\Dout',I')$.
Then $A$ and $A'$ are \emph{isomorphic}, written $A \iso A'$, 
if there exist bijections 
$\funct{\bijF}{F}{F'}$, $\funct{\bijP}{P}{P'}$, $\funct{\bijV}{V}{V'}$, 
$\funct{\bijD}{D}{D'}$, $\funct{\bijI}{\Din}{\Din'}$,
$\funct{\bijO}{\Dout}{\Dout'}$, and 
$\funct{\bijB}{\set{0,1}}{\set{0,1}}$ such that:
\begin{itemize}
\item
$\bijF(\ini) = \ini$ and $\bijF(\fin) = \fin$;
\item
for all $v,v' \in V$, 
$(v,v') \in E$ iff $(\bijV(v),\bijV(v')) \in E'$;
\item
for all $v \in V$ with $l(v) \in F$, $\bijF(l(v)) = l'(\bijV(v))$;
\item
for all $v \in V$ with $l(v) \in P$, $\bijP(l(v)) = l'(\bijV(v))$; 
\item
for all $(v,v') \in E$ with $l((v,v'))$ is defined, 
$\bijB(l((v,v'))) = l'((\bijV(v),\bijV(v')))$;
\item
for all $d \in \Din$, $\bijD(I(\ini)(d)) = I'(\ini)(\bijI(d))$;
\item
for all $d \in D$, $\bijO(I(\fin)(d)) = I'(\fin)(\bijD(d))$;
\item
for all $d \in D$ and $f \in \widetilde{F}$, 
$\bijD(I(f)(d)) = I'(\bijF(f))(\bijD(d))$;
\item
for all $d \in D$ and $p \in P$, 
$\bijB(I(p)(d)) = I'(\bijP(p))(\bijD(d))$.
\end{itemize}
\end{udef}

Proto-algorithms may also be considered too concrete in a way not 
covered by isomorphism of proto-algorithms.
This issue is addressed in Section~\ref{sect-algorithm} and leads there
to the introduction of two other equivalence relations.
Although it is intuitive clear what isomorphism of proto-algorithms is,
its precise definition is not easy to memorize.
The equivalence relations that are given in Section~\ref{sect-algorithm} 
may be easier to memorize.

A proto-algorithm could also be defined as a quadruple 
$(D,\Din,\Dout,\overline{G})$ where $\overline{G}$ is a graph that 
differs from a $\Sigma$-algorithm graph in that its vertex labels are 
computable functions from $\Din$ to $D$, $D$ to $\Dout$, $D$ to $D$ or 
$D$ to $\set{0,1}$ instead of function and predicate symbols from 
$\Sigma$.
I consider the definition of a proto-algorithm given earlier more 
insightful because it isolates as much as possible the operations to be 
performed and the conditions to be inspected from its structure.

\section{Algorithmic and Computational Equivalence}
\label{sect-algorithm}

In Section~\ref{sect-proto-algo}, the intuition was given that a 
proto-algorithm $A$ is something that goes through states.
It was informally explained how the state that it is in determines what 
the next state is. 
The algorithmic step function $\astep_A$ that is defined below 
formalizes this.
The computational step function $\cstep_A$ that is also defined below is 
like the algorithmic step function $\astep_A$, but conceals the steps 
that consist of inspecting conditions.

\begin{udef}
Let $A = (\Sigma,G,\cI)$ be a proto-algorithm, where $\Sigma = (F,P)$, 
$G = (V,E,\LBLv,\LBLe,l,r)$, and $\cI = (D,\Din,\Dout,I)$.
Then the \emph{algorithmic step function $\astep_A$ induced by $A$} is 
the unary total function on the set $\Din \union (V \x D) \union \Dout$ 
defined by:
\begin{trivlist}
\item[]
\renewcommand{\arraystretch}{1.2}
\hspace*{.5em}
\begin{tabular}{@{}l@{}l@{\;}l@{}} 
$\astep_A(d)$ & ${} = (v',d')$ &
if $d \in \Din$,\,\, $(r,v') \in E$, and $I(\ini)(d) = d'$;
\\
$\astep_A((v,d))$ & ${} = (v',d')$ &
if $l(v) = o$, $o \in \widetilde{F}$, $(v,v') \in E$, and 
$I(o)(d) = d'$;
\\
$\astep_A((v,d))$ & ${} = (v',d)$ &
if $l(v) = p$, $p \in P$, $(v,v') \in E$, and $I(p)(d) = l((v,v'))$;
\\
$\astep_A((v,d))$ & ${} = d'$ &
if $l(v) = \fin$ and $I(\fin)(d) = d'$;
\\
$\astep_A(d)$ & ${} = d$ &
if $d \in \Dout$;
\end{tabular}
\end{trivlist}
and the \emph{computational step function $\cstep_A$ induced by $A$} is 
the unary total function on the set $\Din \union (V \x D) \union \Dout$ 
defined by:
\begin{trivlist}
\item[]
\renewcommand{\arraystretch}{1.2}
\hspace*{.5em}
\begin{tabular}{@{}l@{}l@{\;}l@{}} 
$\cstep_A(d)$ & ${} = (v',d')$ &
if $d \in \Din$,\,\, $(r,v') \in E$, and $I(\ini)(d) = d'$;
\\
$\cstep_A((v,d))$ & ${} = (v',d')$ &
if $l(v) = o$, $o \in \widetilde{F}$, $(v,v') \in E$, and 
$I(o)(d) = d'$;
\\
$\cstep_A((v,d))$ & ${} = \cstep_A((v',d))$ &
if $l(v) = p$, $p \in P$, $(v,v') \in E$, and $I(p)(d) = l((v,v'))$;
\\
$\cstep_A((v,d))$ & ${} = d'$ &
if $l(v) = \fin$ and $I(\fin)(d) = d'$;
\\
$\cstep_A(d)$ & ${} = d$ &
if $d \in \Dout$.
\end{tabular}
\end{trivlist}
\end{udef}

If a proto-algorithm $A'$ can mimic a proto-algorithm $A$ step-by-step, 
then we say that $A$ is algorithmically simulated by $A'$.
If the steps that consist of inspecting conditions are ignored, then we 
say that $A$ is computationally simulated by~$A'$.
Algorithmic and computational simulation can be formally defined using 
the step functions defined above.

\begin{udef}
Let $A = (\Sigma,G,\cI)$ and $A' = (\Sigma',G',\cI')$ be two 
proto-algorithms, where
$G = (V,E,\LBLv,\LBLe,l,r)$, $G' = (V',E',\LBLv',\LBLe',l',r')$, 
$\cI = (D,\Din,\Dout,I)$, and $\cI' = (D',\Din',\Dout',I')$.
Then an \emph{algorithmic simulation of $A$ by $A'$} is a set
$R \subseteq 
 (\Din \x \Din') \union ((V \x D) \x (V' \x D')) \union 
 (\Dout \x \Dout')$ 
such that: 
\begin{trivlist}
\item[]
\renewcommand{\arraystretch}{1.2}
\hspace*{1.5em}
\begin{tabular}{@{}l@{\;}l@{}}
if $d \in \Din$,   & then there exists a unique $d' \in \Din'$ such that
$(d,d') \in R$;
\\
if $d' \in \Dout'$, & then there exists a unique $d \in \Dout$ such that
$(d,d') \in R$;
\\
if $(d,d') \in R$, & then $(\astep_A(d),\astep_{A'}(d')) \in R$;
\end{tabular}
\end{trivlist}
and a \emph{computational simulation of $A$ by $A'$} is a set
$R \subseteq 
 (\Din \x \Din') \union ((V \x\nolinebreak D) \x (V' \x D')) \union
 (\Dout \x \Dout')$ 
such that: 
\begin{trivlist}
\item[]
\renewcommand{\arraystretch}{1.2}
\hspace*{1.5em}
\begin{tabular}{@{}l@{\;}l@{}}
if $d \in \Din$,   & then there exists a unique $d' \in \Din'$ such that
$(d,d') \in R$;
\\
if $d' \in \Dout'$, & then there exists a unique $d \in \Dout$ such that
$(d,d') \in R$;
\\
if $(d,d') \in R$, & then $(\cstep_A(d),\cstep_{A'}(d')) \in R$.
\end{tabular}
\end{trivlist}
$A$ \emph{is algorithmically simulated by} $A'$, written $A \asim A'$,
if there exists an algorithmic simulation of $A$ by $A'$. 
\\ 
$A$ \emph{is computationally simulated by} $A'$, written $A \csim A'$, 
if there exists a computational simulation of $A$ by $A'$.
\\ 
$A$ \emph{is algorithmically equivalent to} $A'$, written $A \aeqv A'$, 
if $A \asim A'$ and $A' \asim A$. 
\\
$A$ \emph{is computationally equivalent to} $A'$, written $A \ceqv A'$, 
if $A \csim A'$ and $A' \csim A$.
\end{udef}

The following theorem tells us how isomorphism, algorithmic equivalence, 
and computational equivalence are related.
\begin{theorem}
\label{theorem-equivs}
Let $A$ and $A'$ be proto-algorithms.
Then: 
\begin{trivlist}
\item[]
$\qquad$ (1) $\;$ $A \iso A'$ only if $A \aeqv A'$ 
$\qquad$ (2) $\;$ $A \aeqv A'$ only if $A \ceqv A'$.
\end{trivlist}
\end{theorem}
\begin{proof}
Let $A = (\Sigma,G,\cI)$ and $A' = (\Sigma',G',\cI')$ be proto-algorithms, 
where 
$\Sigma = \nolinebreak (F,P)$, $\Sigma' = (F',P')$, 
$G = (V,E,\LBLv,\LBLe,l,r)$, $G' = (V',E',\LBLv',\LBLe',l',r')$, 
$\cI = (D,\Din,\Dout,I)$, and $\cI' = (D',\Din',\Dout',I')$.

Part~1.
Let $\bijV$, $\bijD$, $\bijI$, and $\bijO$ be as in the definition of 
$\iso$, and
let $\beta$ be the bijection from $\Din \union (V \x D) \union \Dout$ to 
$\Din' \union (V' \x D') \union \Dout'$ defined by:
$\beta(d) = \bijI(d)$ if $d \in \Din$,
$\beta((v,d)) = (\bijV(v),\bijD(d))$, and 
$\beta(d) = \bijO(d)$ if $d \in \Dout$.
It is easy to show that, 
for all $d \in \Din \union (V \x D) \union \Dout$, 
$\beta({\astep_A}(d)) = {\astep_{A'}}(\beta(d))$.
It immediately follows that the set
$\set{(\astepn{n}_A(d),\beta(\astepn{n}_A(d))) \where
      d \in \Din \Land n \in \Nat}$ 
is an algorithmic simulation of $A$ by $A'$.%
\footnote
{The notation $\astepn{n}_A(d)$, where $n \in \Nat$, is used for the 
 $n$-fold application of $\astep_A$ to $d$,
 i.e.\ $\astepn{0}_A(d) = d$ and
 $\astepn{n+1}_A(d) = \astep(\astepn{n}_A(d))$.}
Hence, $A \asim A'$.
The proof of $A' \asim A$ is done in the same way.

Part~2.
Because $A \aeqv A'$, there exists an algorithmic simulation of $A$ 
by~$A'$.
Let $R$ be an algorithmic simulation of $A$ by $A'$.
Then it is easy to show that, for all $(d,d') \in R$, 
$(\cstep_A(d),\cstep_{A'}(d')) \in R$.
It immediately follows that $R$ is also a computational simulation of 
$A$ by $A'$.
Hence, $A \csim A'$.
The proof of $A' \csim A$ is done in the same way.
\qed
\end{proof}

We do not have that $A \iso A'$ if $A \aeqv A'$.
The following example illustrates this.
Take proto-algorithms $A = (\Sigma,G,\cI)$ and $A' = (\Sigma,G',\cI)$ 
where:
\begin{itemize}
\item
$G$ contains edges $(v_1,v_1')$, $(v_1',v'')$, $(v_2,v_2')$, and 
$(v_2',v'')$ where the vertices $v_1'$ and $v_2'$ are labeled by the 
same function symbol;
\item
$G'$ is obtained from $G$ by replacing the edge $(v_2,v_2')$ by 
$(v_2,v_1')$ and removing the edge $(v_2',v'')$.
\end{itemize}
Clearly, $A$ and $A'$ are algorithmically equivalent, but not 
isomorphic.

We also do not have $A \aeqv A'$ if $A \ceqv A'$.
The following example illustrates this.
Take proto-algorithms $A = (\Sigma,G,\cI)$ and $A' = (\Sigma,G',\cI)$ 
where:
\begin{itemize}
\item
$G$ contains a cycle in which only one vertex occurs that is labeled by 
a predicate symbol $p$ and the outgoing edge of this vertex that is not 
part of the cycle is labeled by $1$;
\item
$G'$ is obtained from $G$ by adding immediately before the cycle a copy 
of the cycle in which the predicate symbol $p$ is replaced by a 
predicate symbol $p'$ whose interpretation yields $1$ whenever the 
interpretation of $p$ yields $1$.
\end{itemize}
It is easy to see that $A$ and $A'$ are computationally equivalent, but 
not algorithmically equivalent.

The definition of algorithmic equivalence suggests that the patterns of 
behaviour expressed by algorithmically equivalent proto-algorithms must 
be considered the same.
This suggests in turn that algorithms are equivalence classes of 
proto-algorithms under algorithmic equivalence.

If two proto-algorithms are computationally equivalent, then, for each 
input value, they lead to the same sequence of operations being 
performed. 
The point of view should not be taken that the patterns of 
behaviour expressed by computationally equivalent proto-algorithms are 
the same: the steps that consist of inspecting a condition are treated 
as if they do not belong to the patterns of behaviour.

The relevance of the computational equivalence relation is that any 
equivalence relation that captures the sameness of the patterns of 
behaviour expressed by proto-algorithms to a higher degree than the 
algorithmic equivalence relation must be finer than the computational 
equivalence relation.

\begin{udef}
Let $A = (\Sigma,G,\cI)$ be a proto-algorithm, where $\Sigma = (F,P)$, 
$G = (V,E,\LBLv,\LBLe,l,r)$, and $\cI = (D,\Din,\Dout,I)$.
Then the \emph{function $\widehat{A}$ computed by $A$} is the partial 
function from $\Din$ to $\Dout$ defined by 
$\widehat{A}(d) = \astepr_A(d)$, where $\astepr_A$ is the least-defined 
unary partial function on $\Din \union (V \x D) \union \Dout$ satisfying
\begin{trivlist}
\item[]
\renewcommand{\arraystretch}{1.2}
\hspace*{1.5em}
\begin{tabular}{@{}l@{}l@{\;}l@{}} 
$\astepr_A(d)$ & ${} = \astepr_A(\astep_A(d))$ &
if $\astep_A(d) \in V \x D$;
\\
$\astepr_A(d)$ & ${} = \astep_A(d)$ & 
if $\astep_A(d) \in \Dout$.
\end{tabular}
\end{trivlist} 
Let, moreover, $d \in \Din$ be such that $\widehat{A}(d)$ is defined.
Then the 
\emph{number of algo\-rithmic steps to compute $\widehat{A}(d)$ by $A$},
written $\nas(A,d)$, is the smallest $n \in \Nat$ such that  
$\astepn{n}_A(d) = \widehat{A}(d)$. 
\end{udef}

The following theorem tells us that, 
if a proto-algorithm $A$ is simulated by a proto-algorithm $A'$, then 
(a)~the function computed by $A'$ models the function computed by $A$ 
(in the sense of e.g.~\cite{Jon90a}) and
(b)~for each input value for which $A$ eventually outputs an output 
value, $A'$ does so in the same number of algorithmic steps.
\begin{theorem}
\label{theorem-alg-equiv}
Let $A = (\Sigma,G,\cI)$ and $A' = (\Sigma',G',\cI')$ be 
proto-algorithms, where $\cI = (D,\Din,\Dout,I)$ and 
$\cI' = (D',\Din',\Dout',I')$.
Then $A \asim A'$ only if there exist total functions 
$\funct{\fncI}{\Din}{\Din'}$ and $\funct{\fncO}{\Dout'}{\Dout}$ 
such that: 
\begin{enumerate}
\item[(1)]
for all $d \in \Din$,
$\widehat{A}(d)$ is defined only if $\widehat{A'}(\fncI(d))$ is defined;
\item[(2)]
for all $d \in \Din$ and $d' \in \Dout$,
$\widehat{A}(d) = d'$ only if $\fncO(\widehat{A'}(\fncI(d))) = d'$;
\item[(3)]
for all $d \in \Din$ such that $\widehat{A}(d)$ is defined,
$\nas(A,d) = \nas(A',\fncI(d))$.
\end{enumerate}
\end{theorem}
\begin{proof}
Because $A \asim A'$, there exists an algorithmic simulation of $A$ by 
$A'$.
Let $R$ be an algorithmic simulation of $A$ by $A'$,
let $\fncI$ be the unique function from $\Din$ to $\Din'$ such that, 
for all $d \in \Din$, $(d,\fncI(d)) \in R$, and
let $\fncO$ be the unique function from $\Dout'$ to $\Dout$ such that, 
for all $d' \in \Dout'$, $(\fncO(d'),d') \in R$.
From the definition of an algorithmic simulation, it follows immediately
that, for all $d \in \Din$, for all $n \in \Nat$, 
$(\astepn{n}_A(d),\astepn{n}_{A'}(\fncI(d))) \in R$.
From this result and the definition of an algorithmic simulation, it 
follows immediately that, for all $d \in \Din$, for all $n \in \Nat$:
\begin{enumerate} 
\item[(a)]
$\astepn{n}_A(d) \in \Dout$ iff 
$\astepn{n}_{A'}(\fncI(d)) \in \Dout'$; 
\item[(b)]
for all $d' \in \Dout$, 
$\astepn{n}_A(d) = d'$ iff 
there exists a $d'' \in \Dout'$ such that 
$\astepn{n}_{A'}(\fncI(d)) = d''$ and $\fncO(d'') = d'$.
\end{enumerate}

By the definition of the function computed by a proto-algorithm, we have 
that \linebreak[2]
$\widehat{A}(d)$ is defined iff there exists an $n \in \Nat$ 
such that $\astepn{n}_A(d) \in \Dout$ and that
$\widehat{A'}(\fncI(d))$ is defined iff there exists an $n \in \Nat$ 
such that $\astepn{n}_{A'}(\fncI(d)) \in \Dout'$.
From this and~(a), (1)~follows immediately. 

By the definition of the function computed by a proto-algorithm, we have 
that 
$\widehat{A}(d) = d'$ iff there exists an $n \in \Nat$ 
such that $\astepn{n}_A(d) = d'$ and that
$\widehat{A'}(\fncI(d)) = d''$ iff there exists an $n \in \Nat$ 
such that $\astepn{n}_{A'}(\fncI(d)) = d''$.
From this and~(b), (2)~follows immediately. 

By the definition of $\nas$ and~(a), (3)~also follows immediately.
\qed
\end{proof}
It is easy to see that, for all $d \in \Din$, 
$\widehat{A}(d) = \cstepr_A(d)$, where $\cstepr_A$ is the least-defined 
unary partial function on $\Din \union (V \x D) \union \Dout$ satisfying
\begin{trivlist}
\item[]
\renewcommand{\arraystretch}{1.2}
\hspace*{1.5em}
\begin{tabular}{@{}l@{}l@{\;}l@{}} 
$\cstepr_A(d)$ & ${} = \cstepr_A(\cstep_A(d))$ &
if $\cstep_A(d) \in V \x D$;
\\
$\cstepr_A(d)$ & ${} = \cstep_A(d)$ & 
if $\cstep_A(d) \in \Dout$.
\end{tabular}
\end{trivlist}
This means that Theorem~\ref{theorem-alg-equiv} goes through as far as 
(1) and~(2) are concerned if algorithmic simulation is replaced by 
computational simulation.
It follows immediately from the example of computationally equivalent 
proto-algorithms given earlier that (3) does not go through if 
algorithmic simulation is replaced by computational simulation.

\section{The Imperative Process Algebra \deBPAde}
\label{sect-deBPAde}

In Section~\ref{sect-algo-procs}, a connection is made between  
proto-algorithms and the processes that are considered in the imperative 
process algebra \deBPAde.
In this section, a short survey of \deBPAde\ and recursion in the 
setting of \deBPAde\ is given.
The constants and operators of the algebraic theory \deBPAde\ and the
additional constants of its extension with recursion are discussed.
The axioms of \deBPAde\ are given in the Appendix.
\deBPAde\ is a subtheory of \deACPet.
In~\cite{Mid21a}, a comprehensive treatment of \deACPet\ can be found.
The axioms of \deBPAde\ are the axioms of \deACPet\ in which only 
constants and operators of \deBPAde\ occur.
The additional axioms of the extension of \deBPAde\ with recursion are
simply the additional axioms of the extension of \deACPet\ with 
recursion.

\subsection{\BPA\ with Inaction and Empty Process}
\label{subsect-BPAde}

First, a short survey of \BPAde\ is given.
\BPAde\ is the version of \BPA\ with inaction and empty process constants 
that was first presented in~\cite[Section~2.2]{BW90}.
In Section~\ref{subsect-deBPAde}, \deBPAde\ will be introduced as an 
extension of \BPAde.

In \BPAde, it is assumed that a fixed but arbitrary finite set $\Act$ of 
\emph{basic actions}, with $\dead,\ep \not\in \Act$, has been given.
Basic actions are taken as atomic processes.

The algebraic theory \BPAde\ has one sort: the sort $\Proc$ of 
\emph{processes}.
This sort is made explicit to anticipate the need for many-sortedness 
later on. 
The algebraic theory \BPAde\ has the following constants and operators 
to build terms of sort~$\Proc$:
\begin{itemize}
\item
a \emph{basic action} constant $\const{a}{\Proc}$ for each 
$a \in \Act$;
\item
an \emph{inaction} constant $\const{\dead}{\Proc}$;
\item
an \emph{empty process} constant $\const{\ep}{\Proc}$;
\item
a binary \emph{alternative composition} or \emph{choice} operator 
$\funct{\altc}{\Proc \x \Proc}{\Proc}$;
\item
a binary \emph{sequential composition} operator 
$\funct{\seqc}{\Proc \x \Proc}{\Proc}$.
\end{itemize}
It is assumed that there is a countably infinite set $\cX$ of variables 
of sort $\Proc$, which contains $x$, $y$ and $z$.
Terms are built as usual.
Infix notation is used for the operators ${} \altc {}$ and 
${} \seqc {}$.
The following precedence convention are used to reduce the need for
parentheses: the operator ${} \seqc {}$ binds stronger than the operator 
${} \altc {}$.

The constants $a$ ($a \in \Act$), $\ep$, and $\dead$ can be 
explained as follows:
(a)~$a$ denotes the process that first performs the action $a$ and then 
terminates successfully,
(b)~$\ep$ denotes the process that terminates successfully without 
performing any action, and
(c)~$\dead$ denotes the process that cannot do anything, it cannot even 
terminate successfully.

Let $t$ and $t'$ be closed \BPAde\ terms. 
Then the operators $\altc$ and $\seqc$ can be explained as follows:
(a)~$t \altc t'$ denotes the process that behaves as the process 
denoted by $t$ or as the process denoted by $t'$, where the choice 
between the two is resolved at the instant that one of them does 
something, and
(b)~$t \seqc t'$ denotes the process that first behaves as 
the process denoted by $t$ and following successful termination of that 
process behaves as the process denoted by $t'$.

\subsection{Imperative \BPAde}
\label{subsect-deBPAde}

\deBPAde, imperative \BPAde, extends \BPAde\ with features to change 
data involved in a process in the course of the process and to proceed 
at certain stages of a process in a way that depends on the changing 
data. 

In \deBPAde, it is assumed that the following has been given with 
respect to data:
\begin{itemize}
\item
a many-sorted signature $\sign_\gD$ that includes:
\begin{itemize}
\item
a sort $\Data$ of \emph{data} and
a sort $\Bool$ of \emph{bits};
\item
constants of sort $\Data$ and/or operators with result sort $\Data$;
\item
constants $\zero$ and $\one$ of sort $\Bool$ and
operators with result sort $\Bool$;
\end{itemize}
\item
a minimal algebra $\gD$ of signature $\sign_\gD$ in which the carrier of 
sort $\Bool$ has cardinality $2$ and the equation $\zero = \one$ does 
not hold.
\end{itemize}
We write $\DataVal$ for the set of all closed terms over the signature
$\sign_\gD$ of sort $\Data$.

In \deBPAde, it is moreover assumed that a finite or countably infinite 
set $\FlexVar$ of \emph{flexible variables} has been given.
A flexible variable is a variable whose value may change in the course 
of a process.%
\footnote
{The term flexible variable is used for this kind of variables in 
 e.g.~\cite{Lam94a,Sch97a}.} 

A \emph{flexible variable valuation} is a total function from $\FlexVar$ 
to $\DataVal$. 
We write $\FVarVal$ for the set of all flexible variable valuations.

Flexible variable valuations provide closed terms from $\DataVal$ that 
denote the members of $\gD$'s carrier of sort $\Data$ assigned to 
flexible variables when a \deBPAde\ term of sort $\Data$ is evaluated.
Because $\gD$ is a minimal algebra, each member of $\gD$'s carrier of 
sort $\Data$ can be represented by a term from $\DataVal$. 
We write $d$, where $d$ is a member of $\gD$'s carrier of sort 
$\Data$, for a fixed but arbitrary term from $\DataVal$ representing $d$ 
when it is clear from the context that a term from $\DataVal$ is 
expected.

\deBPAde\ has the following sorts: 
the sorts included in $\sign_\gD$,
the sort $\Cond$ of \emph{conditions}, and
the sort $\Proc$ of \emph{processes}.

For each sort $s$ included in $\sign_\gD$ other than $\Data$, 
\deBPAde\ has only the constants and operators included in $\sign_\gD$ 
to build terms of sort $s$.

\deBPAde\ has, in addition to the constants and operators included in 
$\sign_\gD$ to build terms of sorts $\Data$, the following constants to 
build terms of sort $\Data$:
\begin{itemize}
\item
for each $v \in \FlexVar$, the \emph{flexible variable} constant 
$\const{v}{\Data}$.
\end{itemize}
We write $\DataTerm$ for the set of all closed \deBPAde\ terms of sort 
$\Data$.

\deBPAde\ has the following constants and operators to build terms of 
sort~$\Cond$:
\begin{itemize}
\item
a binary \emph{equality} operator
$\funct{\Leq}{\Bool \x \Bool}{\Cond}$;
\item
a binary \emph{equality} operator
$\funct{\Leq}{\Data \x \Data}{\Cond}$;%
\footnote
{The overloading of $=$ can be trivially resolved if $\sign_\gD$ is
 without overloaded symbols.}
\item
a \emph{truth} constant $\const{\True}{\Cond}$;
\item
a \emph{falsity} constant $\const{\False}{\Cond}$;
\item
a unary \emph{negation} operator $\funct{\Lnot}{\Cond}{\Cond}$;
\item
a binary \emph{conjunction} operator 
$\funct{\Land}{\Cond \x \Cond}{\Cond}$;
\item
a binary \emph{disjunction} operator 
$\funct{\Lor}{\Cond \x \Cond}{\Cond}$;
\item
a binary \emph{implication} operator 
$\funct{\Limpl}{\Cond \x \Cond}{\Cond}$;
\end{itemize}
We write $\CondTerm$ for the set of all closed \deBPAde\ terms of sort 
$\Cond$.

\deBPAde\ has, in addition to the constants and operators of \BPAde, 
the following operators to build terms of sort $\Proc$:
\begin{itemize}
\item
a unary \emph{assignment action} operator
$\funct{\assop{v}\,}{\Data}{\Proc}$ for each $v \in \FlexVar$;
\item
a binary \emph{guarded command} operator 
$\funct{\gc\,}{\Cond \x \Proc}{\Proc}$;
\item
a unary \emph{evaluation} operator 
$\funct{\eval{\rho}}{\Proc}{\Proc}$ for each $\rho \in \FVarVal$.
\end{itemize}
We write $\ProcTerm$ for the set of all closed \deBPAde\ terms of sort 
$\Proc$.

It is assumed that there are countably infinite sets of variables of 
sort $\Data$ and $\Cond$ and that the sets of variables of sort $\Data$, 
$\Cond$, and $\Proc$ are mutually disjoint and disjoint from $\FlexVar$.

The same notational conventions are used as before.
Infix notation is also used for the additional binary operators.
Moreover, the notation $\ass{v}{e}$, where $v \in \FlexVar$ and $e$ is a 
\deBPAde\ term of sort $\Data$, is used for the term $\assop{v}(e)$.

Each term from $\CondTerm$ can be taken as a formula of a first-order 
language with equality of $\gD$ by taking the flexible variable
constants as variables of sort $\Data$.
The flexible variable constants are implicitly taken as variables of 
sort $\Data$ wherever the context asks for a formula.
In this way, each term from $\CondTerm$ can be interpreted in $\gD$ as a
formula.

The notation $\phi \Liff \psi$, where $\phi$ and $\psi$ are 
\deBPAde\ terms of sort $\Cond$, is used for the term
$(\phi \Limpl \psi) \Land (\psi \Limpl \phi)$.
The axioms of \deBPAde\ include an equation $\phi = \psi$ for each two 
terms $\phi$ and $\psi$ from $\CondTerm$ for which the formula 
$\phi \Liff \psi$ holds in $\gD$.

Let 
$e$ be a term from $\DataTerm$, $\phi$ be a term from $\CondTerm$, and 
$t$ be a term from $\ProcTerm$.
Then the additional operators to build terms of sort $\Proc$ can be 
explained as follows:
\begin{itemize}
\item
the term $\ass{v}{e}$ denotes the process that first performs the 
assignment action $\ass{v}{e}$, whose intended effect is the assignment 
of the result of evaluating $e$ to flexible variable $v$, and then 
terminates successfully; 
\item
the term $\phi \gc t$ denotes the process that behaves as the process 
denoted by $t$ if condition $\phi$ holds and as $\dead$ otherwise;
\item
the term $\eval{\rho}(t)$ denotes the process that behaves as the process
denoted by $t$, except that each subterm of $t$ that belongs to 
$\DataTerm$ is evaluated using flexible variable valuation $\rho$ updated 
according to the assignment actions that have taken place at the point 
where the subterm is encountered.
\end{itemize}

Below will be referred to the subset $\AProcTerm$ of $\ProcTerm$ that 
consists of the terms from $\ProcTerm$ that denote the processes that
are considered to be atomic.

$\AProcTerm$ is defined as follows:
$\AProcTerm  =   
 \Act \union
 \set{\ass{v}{e} \where v \in \FlexVar \Land e \in \DataTerm}$.

\subsection{\deBPAde\ with Recursion}
\label{subsect-deBPAder}

In this section, recursion in the setting of \deBPAde\ is treated.
A closed \deBPAde\ term of sort $\Proc$ denotes a process with a finite 
upper bound to the number of actions that it can perform. 
Recursion allows the description of processes without a finite upper 
bound to the number of actions that it can perform.

A \emph{recursive specification} over \deBPAde\ is a set 
$\set{X = t_X \where X \in V}$ of \emph{recursion equations}, where 
$V$ is a subset of $\cX$ and each $t_X$ is a \deBPAde\ term of sort 
$\Proc$ in which only variables from $V$ occur. 
We write $\vars(S)$, where $S$ is a recursive specification over 
\deBPAde, for the set of all variables that occur in $S$.

A \emph{solution} of a recursive specification $S$ over \deBPAde\ in 
some model of \deBPAde\ is a set $\set{p_X \where X \in \vars(S)}$ of 
elements of the carrier of sort $\Proc$ in that model such that each 
equation in $S$ holds if, for all $X \in \vars(S)$, $X$ is assigned 
$p_X$. 
If $\set{p_X \where X \in \vars(S)}$ is a solution of a recursive 
specification $S$, then, for each $X \in \vars(S)$, $p_X$ is called the 
\emph{$X$-component} of that solution of $S$. 
Each recursive specification over \deBPAde\ that has a unique solution 
in the model of \deBPAde\ given in~\cite{Mid21a} can be rewritten to a 
recursive specification in which the right-hand sides of equations are 
linear \deBPAde\ terms.

The set $\LT$ of \emph{linear \deBPAde\ terms} is inductively defined by 
the following~rules:
\begin{itemize}
\item
$\dead \in \LT$;
\item 
if $\phi \in \CondTerm$, then $\phi \gc \ep \in \LT$;
\item
if $\phi \in \CondTerm$, $\alpha \in \AProcTerm$, and $X \in \cX$, then 
$\phi \gc \alpha \seqc X \in \LT$;
\item
if $t,t' \in \LT \diff \set{\dead}$, then $t \altc t' \in \LT$.
\end{itemize}

A \emph{linear recursive specification} over \deBPAde\ is a recursive 
specification $\set{X = \nolinebreak t_X \where X \in V}$ over \deBPAde\ 
where each $t_X \in \LT$.

\deBPAde\ is extended with recursion by adding constants for solutions 
of linear recursive specifications over \deBPAde\ and axioms concerning 
these additional constants.
For each linear recursive specification $S$ over \deBPAde\ and 
each $X \in \vars(S)$, a constant $\rec{X}{S}$ of sort $\Proc$ is added 
to the constants of \deBPAde\ and axioms postulating that $\rec{X}{S}$
stands for the $X$-component of the unique solution of $S$ are added to
the axioms of \deBPAde.
We write \deBPAder\ for the resulting theory.

We write $\ProcTermr$ for the set of all closed $\deBPAder$ terms of 
sort $\Proc$.
We write ${} \Ent t = t'$, where $t$ and $t'$ are $\deBPAder$ terms of 
sort $\Proc$, to indicate that the equation $t = t'$ is derivable from
the axioms of $\deBPAder$.

\section{Algorithm Processes}
\label{sect-algo-procs}

In this section, a connection is made between proto-algorithms and the
processes considered in the imperative process algebra \deBPAde.
It is assumed that $\MEM \in \FlexVar$.

\begin{udef}
Let $\Sigma = (F,P)$ be an alphabet.
Then a \emph{$\Sigma$-algorithm process} is a constant $\rec{X}{S}$ of 
\deBPAde, where $S$ is finite, $X_\ep \in \vars(S)$, and for each 
$Y \in \vars(S)$:
\begin{itemize}
\item
the recursion equation for $Y$ in $S$ has one of the following forms:
\begin{trivlist}
\item[]
\renewcommand{\arraystretch}{1.2}
\hspace*{1.5em}
\begin{tabular}[t]{@{}l@{\;\;}l@{}}
(1) & $Y = \True \gc \ass{\MEM}{\ini(\MEM)} \seqc Z$, 
\\
(2) & $Y = \True \gc \ass{\MEM}{o(\MEM)} \seqc Z$, 
\\
(3) & $Y = (p(\MEM) = 1) \gc \ass{\MEM}{\MEM} \seqc Z \altc 
     (p(\MEM) = 0) \gc \ass{\MEM}{\MEM} \seqc Z'$, 
\\
(4) & $Y = \True \gc \ass{\MEM}{\fin(\MEM)} \seqc X_\ep$, 
\\
(5) & $Y = \True \gc \ep$,
\end{tabular}
\end{trivlist}%
where $o \in \widetilde{F}$, $p \in P$, and 
$Z,Z' \in \vars(S) \diff \set{X_\ep}$;
\item
the recursion equation for $Y$ in $S$ is of the form~(1) iff 
$Y \equiv X$;
\item
the recursion equation for $Y$ in $S$ is of the form~(5) iff 
$Y \equiv X_\ep$.
\end{itemize} 
\end{udef}
We write $\AlgGr{\Sigma}$ and $\AlgPr{\Sigma}$, where $\Sigma$ is an 
alphabet, for the set of all $\Sigma$-algorithm graphs and the set of 
all $\Sigma$-algorithm processes, respectively.
\begin{udef}
Let $\Sigma = (F,P)$ be an alphabet. 
Then the \emph{graph-to-process} function $\grtopr{\Sigma}$ is a total
function from $\AlgGr{\Sigma}$ to $\AlgPr{\Sigma}$ such that, 
for each $\Sigma$-algorithm graph $G = (V,E,\LBLv,\LBLe,l,r)$, 
$\grtopr{\Sigma}(G) = \rec{X}{S}$, where $\rec{X}{S}$ is a 
$\Sigma$-algorithm process such that:
\begin{trivlist}
\item[]
\renewcommand{\arraystretch}{1.2}
\hspace*{.5em}
\begin{tabular}{@{}l@{}l@{\;}l@{}} 
$X$   & ${} = \True \gc \ass{\MEM}{\ini(\MEM)} \seqc X_{v'} \,\in\, S$ &
iff $(r,v') \in E$; 
\\
$X_v$ & ${} = \True \gc \ass{\MEM}{o(\MEM)} \seqc X_{v'} \,\in\, S$ &
iff $v \in V$, $l(v) = o$, $o \in \widetilde{F}$, $(v,v') \in E$;
\\
$X_v$ & \multicolumn{2}{@{}l@{}}
       {${} =
 p(\MEM) = 1 \gc \ass{\MEM}{\MEM} \seqc X_{v'} \altc
 p(\MEM) = 0 \gc \ass{\MEM}{\MEM} \seqc X_{v''} \,\in\, S$} \\ 
 \multicolumn{3}{@{\quad}l@{}}
       {
 iff $v \in V$, $l(v) = p$, $p \in P$, $(v,v'),(v,v'') \in E$, 
 $l((v,v')) = 1$, $l((v,v'')) = 0$;}
\\
$X_v$   & ${} = \True \gc \ass{\MEM}{\fin(\MEM)} \seqc X_\ep \,\in\, S$ &
iff $v \in V$, $l(v) = \fin$; 
\\
$X_\ep$ & ${} = \True \gc \ep$;
\end{tabular}
\end{trivlist}
where, for all $v \in V$, $X_v \in \cX$ and, for all $v' \in V$, 
$X_v = X_{v'}$ only if $v = v'$.
\end{udef}
The function $\grtopr{\Sigma}$ is uniquely defined up to renaming of
variables.

The following theorem tells us that the function $\grtopr{\Sigma}$ is a 
bijection from $\AlgGr{\Sigma}$ to $\AlgPr{\Sigma}$ up to isomorphism of 
algorithm graphs and renaming of variables in algorithm processes.
\begin{theorem}
\label{theorem-grtopr}
Let $\Sigma$ be an alphabet.
Then, for all $\rec{X}{S} \in \AlgPr{\Sigma}$, there exists a unique 
$G \in \AlgGr{\Sigma}$ up to $\iso$ such that $\rec{X}{S}$ and 
$\grtopr{\Sigma}(G)$ are identical up to consistent renaming of 
variables.
\end{theorem}
\begin{proof}
Let $\Sigma = (F,P)$ be an alphabet, and 
let $\rec{X}{S} \in \AlgPr{\Sigma}$. 
Then we construct a $G = (V,E,\LBLv,\LBLe,l,r) \in \AlgGr{\Sigma}$ from
$\rec{X}{S}$ as follows:
\begin{itemize}
\item
$V = vars(S) \diff \set{X_\ep}$;
\item
$E$ is the set of all $(Y,Z) \in V \x V$ for which there exists an 
equation in $S$ such that $Y$ is its left-hand side and $Z$ occurs in 
its right-hand side;
\item
$\LBLv = F \union P$;
\item
$\LBLe = \set{0,1}$;
\item
$l$ is defined as follows:
\begin{list}{$\bullet$}{\leftmargin=.8em}
\item
$l(X) = \ini$;
\item
$l(Y) = o$ if  
$Y = \True \gc \ass{\MEM}{o(\MEM)} \seqc Z \in S$ 
for some $Z \in \vars(S)$;
\item
$l(Y) = p$ if
$Y = (p(\MEM) = 1) \gc \ass{\MEM}{\MEM} \seqc Z \altc 
                  (p(\MEM) = 0) \gc \ass{\MEM}{\MEM} \seqc Z' \in S$ 
for some $Z,Z' \in \vars(S)$;
\item
$l(Y) = \fin$ if 
$Y = \True \gc \ass{\MEM}{\fin(\MEM)} \seqc X_\ep \in S$;
\item
$l((Y,Z))$ is undefined if 
$Y = \True \gc \ass{\MEM}{o(\MEM)} \seqc Z \in S$ 
for some $o \in F$; 
\item
$l((Y,Z)) = 1$ if  
$Y = (p(\MEM) = 1) \gc \ass{\MEM}{\MEM} \seqc Z \altc 
     (p(\MEM) = 0) \gc \ass{\MEM}{\MEM} \seqc Z' \in S$ 
for some $p \in P$ and $Z' \in \vars(S)$;
\item
$l((Y,Z)) = 0$ if 
$Y = (p(\MEM) = 1) \gc \ass{\MEM}{\MEM} \seqc Z' \altc 
     (p(\MEM) = 0) \gc \ass{\MEM}{\MEM} \seqc Z \in S$ 
for some $p \in P$ and $Z' \in \vars(S)$;
\end{list}
\item
$r = X$.
\end{itemize}
It is easy to see that $\grtopr{\Sigma}(G)$ and $\rec{X}{S}$ are 
identical up to consistent renaming of variables and that, 
for all $G' \in \AlgGr{\Sigma}$, $\grtopr{\Sigma}(G')$ and $\rec{X}{S}$ 
are identical up to consistent renaming of variables only if $G' \iso G$.
\qed
\end{proof}

It is easy to obtain the signature $\sign_\gD$ and the minimal algebra 
$\gD$ of signature $\sign_\gD$ for a given alphabet $\Sigma$ and a given 
$\Sigma$-interpretation $(D,\Din,\Dout,I)$ after the following issues 
have been addressed:
(a)~$D \union \Din \union \Dout$ must be taken as $\gD$'s carrier of 
sort $\Data$ and consequently the interpretation of each symbol from 
$\Sigma$ must be extended to $D \union \Din \union \Dout$ and
(b)~each member of $\Din$ must be representable by a closed term of
sort $\Data$.
Any extension of the functions concerned may be chosen here because we
comply with the convention to use each of them only if it is known that 
the value to which it is applied belongs to its original domain.
For simplicity, we take all members of $\Din$ as constants of 
sort~$\Data$.

Below, we write $[\MEM \mapsto d]$, where $d$ is a member of $\gD$'s carrier of 
sort $\Data$, for a fixed but arbitrary 
$\rho \in \FVarVal$ such that $\rho(\MEM) = d$. 

The graph-to-process function $\grtopr{\Sigma}$ allows to characterize 
the algorithmic step function of a proto-algorithm $A = (\Sigma,G,\cI)$ 
in \deBPAder.
\begin{lemma}
\label{lemma-grtopr}
Let $A = (\Sigma,G,\cI)$ be a proto-algorithm, where 
$\cI = (D,\Din,\Dout,I)$ and 
$G = (V,E,\LBLv,\LBLe,l,r) \in \AlgGr{\Sigma}$, 
let $\rec{X}{S} \in  \AlgPr{\Sigma}$ be such that 
$\rec{X}{S} = \grtopr{\Sigma}(G)$.
Then, for all $v,v_1,v_2 \in V$, $d,d_1,d_2 \in D$, $\din \in \Din$, and 
$\dout \in \Dout$:
\begin{trivlist}
\item[]
\renewcommand{\arraystretch}{1.2}
\hspace*{.1em}
\begin{tabular}{@{}l@{}l@{\;}l@{}l@{}} 
$\astep_A(\din)$ & ${} = (v,d)$   & iff 
${} \Ent \eval{[\MEM \mapsto \din]}(\rec{X}{S})$   & ${} =
 \ass{\MEM}{d} \seqc \eval{[\MEM \mapsto d]}(\rec{X_v}{S})$,
\\
$\astep_A((v_1,d_1))$ & ${} = (v_2,d_2)$ & iff 
${} \Ent \eval{[\MEM \mapsto d_1]}(\rec{X_{v_1}}{S})$ & ${} =
 \ass{\MEM}{d_2} \seqc \eval{[\MEM \mapsto d_2]}(\rec{X_{v_2}}{S})$, 
\\
$\astep_A((v,d))$ & ${} = \dout$ & iff 
${} \Ent \eval{[\MEM \mapsto d]}(\rec{X_v}{S})$ & ${} =
 \ass{\MEM}{\dout} \seqc \eval{[\MEM \mapsto \dout]}(\rec{X_\ep}{S})$. 
\end{tabular}
\end{trivlist}
\end{lemma}
\begin{proof}
This follows easily from the definition of the algorithmic step function 
$\astep_A$, the definition of the graph-to-process function
$\grtopr{\Sigma}$, and the axioms of \deBPAder.
\qed
\end{proof}

There exists a sound method for proving algorithmic equivalence of two
proto-algorithms $A = (\Sigma,G,\cI)$ and $A' = (\Sigma,G',\cI)$ based 
on the graph-to-process function $\grtopr{\Sigma}$.
\begin{theorem}
\label{theorem-grtopr-algeqv}
Let $A = (\Sigma,G,\cI)$ and $A' = (\Sigma,G',\cI)$ be proto-algorithms,
where $\cI = (D,\Din,\Dout,I)$.
Then $A \aeqv A'$ if, for all $d \in \Din$,  
${} \Ent \eval{[\MEM \mapsto d]}(\grtopr{\Sigma}(G)) =
         \eval{[\MEM \mapsto d]}(\grtopr{\Sigma}(G'))$.
\end{theorem}
\begin{proof}
Suppose that 
$G = (V,E,\LBLv,\LBLe,l,r)$ and $G' = (V',E',\LBLv,\LBLe,l',r')$.
Let $\rec{X}{S}, \rec{X}{S'} \in \AlgPr{\Sigma}$ and 
let $\din \in \Din$.

Starting from $\eval{[\MEM \mapsto \din]}(\rec{X}{S})$, either 
there exists an $n \in \Nat$, such that, 
for some $v_1,\ldots,v_{n+1} \in V$, $d_1,\ldots,d_{n+1} \in D$, 
and $\dout \in \Dout$: 
\begin{trivlist}
\item[]
\renewcommand{\arraystretch}{1.2}
\hspace*{1.5em}
\begin{tabular}{@{}l@{}l@{}} 
${} \Ent
 \eval{[\MEM \mapsto \din]}(\rec{X}{S})$ & ${} =
 \ass{\MEM}{d_1} \seqc \eval{[\MEM \mapsto d_1]}(\rec{X_{v_1}}{S})$, 
\\
${} \Ent
 \eval{[\MEM \mapsto d_i]}(\rec{X_{v_i}}{S})$ & ${} =
 \ass{\MEM}{d_{i+1}} \seqc
 \eval{[\MEM \mapsto d_{i+1}]}(\rec{X_{v_{i+1}}}{S})$\phantom{,} \\
\multicolumn{2}{@{}r@{}}
 {for each $i \in \set{1,\ldots,n}$,}  
\\
${} \Ent
 \eval{[\MEM \mapsto d_{n+1}]}(\rec{X_{v_{n+1}}}{S})$ & ${} =
 \ass{\MEM}{\dout} \seqc \eval{[\MEM \mapsto \dout]}(\rec{X_\ep}{S})$
\end{tabular}
\end{trivlist}
or, for some $v_1,v_2,\ldots \in V$ and $d_1,d_2,\ldots \in D$: 
\begin{trivlist}
\item[]
\renewcommand{\arraystretch}{1.2}
\hspace*{1.5em}
\begin{tabular}{@{}l@{}l@{}} 
${} \Ent
 \eval{[\MEM \mapsto \din]}(\rec{X}{S})$ & ${} =
 \ass{\MEM}{d_1} \seqc \eval{[\MEM \mapsto d_1]}(\rec{X_{v_1}}{S})$, 
\\
${} \Ent
 \eval{[\MEM \mapsto d_i]}(\rec{X_{v_i}}{S})$ & ${} =
 \ass{\MEM}{d_{i+1}} \seqc
 \eval{[\MEM \mapsto d_{i+1}]}(\rec{X_{v_{i+1}}}{S})$\phantom{,} \\
\multicolumn{2}{@{}r@{}}
 {\hspace*{2em} for each $i \in \Nat$.}  
\end{tabular}
\end{trivlist}
From this, using 
$\Ent \eval{[\MEM \mapsto \din]}(\grtopr{\Sigma}(G)) =
         \eval{[\MEM \mapsto \din]}(\grtopr{\Sigma}(G'))$
and the fact that $\Ent \alpha \seqc t = \alpha' \seqc t'$ 
(where $\alpha,\alpha' \in \AProcTerm$ and $t,t' \in \ProcTermr$) 
only if ${} \Ent \alpha = \alpha'$ and ${} \Ent t = t'$, it follows by 
an inductive argument that 
(for $i \in \set{1,\ldots,n}$ or $i \in \Nat$):
\begin{trivlist}
\item[]
\renewcommand{\arraystretch}{1.2}
\hspace*{-4.9pt}
\begin{tabular}{@{}l@{}} 
${} \Ent
 \eval{[\MEM \mapsto \din]}(\rec{X}{S}) =
 \ass{\MEM}{d_1} \seqc \eval{[\MEM \mapsto d_1]}(\rec{X_{v_1}}{S})$ 
only if \\
${} \Ent
 \eval{[\MEM \mapsto \din]}(\rec{X}{S'}) =
 \ass{\MEM}{d_1} \seqc \eval{[\MEM \mapsto d_1]}(\rec{X_{v_1'}}{S'})$
for some $v_1' \in V'$, 
\\[1.5ex]
${} \Ent
 \eval{[\MEM \mapsto d_i]}(\rec{X_{v_i}}{S}) =
 \ass{\MEM}{d_{i+1}} \seqc
 \eval{[\MEM \mapsto d_{i+1}]}(\rec{X_{v_{i+1}}}{S})$
only if \\
${} \Ent
 \eval{[\MEM \mapsto d_i]}(\rec{X_{v_i'}}{S'}) =
 \ass{\MEM}{d_{i+1}} \seqc
 \eval{[\MEM \mapsto d_{i+1}]}(\rec{X_{v_{i+1}'}}{S'})$
for some $v_i',v_{i+1}' \in V'$, 
\\[1.5ex]
${} \Ent
 \eval{[\MEM \mapsto d_{n+1}]}(\rec{X_{v_{n+1}}}{S}) =
 \ass{\MEM}{\dout} \seqc \eval{[\MEM \mapsto \dout]}(\rec{X_\ep}{S})$
only if \\
${} \Ent
 \eval{[\MEM \mapsto d_{n+1}]}(\rec{X_{v_{n+1}'}}{S'}) =
 \ass{\MEM}{\dout} \seqc \eval{[\MEM \mapsto \dout]}(\rec{X_\ep}{S'})$
for some $v_{n+1}' \in V'$. 
\end{tabular}
\end{trivlist}
From this, using Lemma~\ref{lemma-grtopr}, it directly follows that
(for $i \in \set{1,\ldots,n}$ or $i \in \Nat$):
\begin{trivlist}
\item[]
\renewcommand{\arraystretch}{1.2}
\hspace*{1.5em}
\begin{tabular}{@{}l@{}} 
$\astep_A(\din) = (v_1,d_1)$ only if 
$\astep_{A'}(\din) = (v_1',d_1)$ for some $v_1' \in V'$,
\\
$\astep_A((v_i,d_i)) = (v_{i+1},d_{i+1})$ only if \\ 
\phantom{$\astep_A(\din) = (v_1,d_1)$ only if}
$\astep_{A'}((v_i',d_i)) = (v_{i+1}',d_{i+1})$
  for some $v_i',v_{i+1}' \in V'$,
\\
$\astep_A((v_{n+1},d_{n+1})) = \dout$ only if \\ 
\phantom{$\astep_A(\din) = (v_1,d_1)$ only if} 
$\astep_A((v_{n+1}',d_{n+1})) = \dout$ for some $v_{n+1}' \in V'$. 
\end{tabular}
\end{trivlist}
This means that there exists an algorithmic simulation of $A$ by $A'$.
In the same way, we can show that there exists an algorithmic simulation 
of $A'$ by $A$.
Hence, $A \aeqv A'$.
\qed
\end{proof}

We do not have that
$A \aeqv A'$ only if, for all $d \in \Din$,  
${} \Ent \eval{[\MEM \mapsto d]}(\grtopr{\Sigma}(G)) =
         \eval{[\MEM \mapsto d]}(\grtopr{\Sigma}(G'))$.
The following example illustrates this.
Take proto-algorithms $A = (\Sigma,G,\cI)$ and $A' = (\Sigma,G',\cI)$,
where 
$\Sigma = (F,P)$, $G = (V,E,\LBLv,\LBLe,l,r)$, $\cI = (D,\Din,\Dout,I)$, 
and $G'$ is obtained from $G$ by interchanging the labels of two 
vertices $v,v' \in V$ for which $(v,v') \in E$, $\indeg(v') = 1$, 
$l(v),l(v') \in F$ and, for all $d \in D$, 
$I(l(v))(I(l(v'))(d)) = I(l(v'))(I(l(v))(d))$.
This means that two steps, the latter of which is always immediately 
preceded by the first, and that consist of performing an operation, 
where the operations in question are independent, are interchanged.
It is easy to see that $A$ and $A'$ are algorithmically equivalent.
However, because of a different order of certain assignment actions 
in $\grtopr{\Sigma}(G)$ and $\grtopr{\Sigma}(G')$, we do not have that
${} \Ent \eval{[\MEM \mapsto d]}(\grtopr{\Sigma}(G)) = 
         \eval{[\MEM \mapsto d]}(\grtopr{\Sigma}(G'))$.

We also do not have that
$A \iso A'$ if, for all $d \in \Din$,  
${} \Ent \eval{[\MEM \mapsto d]}(\grtopr{\Sigma}(G)) =
         \eval{[\MEM \mapsto d]}(\grtopr{\Sigma}(G'))$.
This is illustrated by the same example as the one used to illustrate 
that we do not have that $A \iso A'$ if $A \aeqv A'$.
 
\section{Discussion of Generalizations}
\label{sect-discussion}

The notion of a proto-algorithm introduced in this paper is based on the
classical informal notion of an algorithm.
Several generalizations of that notion have been proposed, e.g.\ the 
notion of a non-deterministic algorithm, the notion of a parallel 
algorithm, and the notion of an interactive algorithm.

The generalization of the notion of a proto-algorithm to a notion of
a non-deterministic proto-algorithm is easy: weaken, in the definition 
of a $\Sigma$-algorithm graph, the outdegree of vertices labeled with a 
function symbol other than $\fin$ to greater than zero.
In the case of a non-deterministic proto-algorithm, most definitions
involving one or more proto-algorithms and the definition of a
$\Sigma$-algorithm process need an obvious adaptation.
However, the definition of algorithmic equivalence needs an adaptation
that is not obvious at first sight: two non-deterministic 
proto-algorithms $A$ and $A'$ are algorithmically equivalent if there
exist an algorithmic simulation $R$ of $A$ by $A'$ and an algorithmic 
simulation $R'$ of $A'$ by $A$ such that $R' = R^{-1}$.
The condition $R' = R^{-1}$ is necessary to guarantee that $A$ and $A'$ 
have the same choice structure.
With this adaptation, Theorem~\ref{theorem-grtopr-algeqv} goes through
for non-deterministic proto-algorithms.

The generalization of the notion of a proto-algorithm to a notion of
a parallel proto-algorithm is not so easy.
The main reason for this is that there is no consensus about what 
properties are essential for a parallel algorithm.
A parallel algorithm is usually informally described by a sentence like
``A parallel algorithm is an algorithm in which more than one step can 
take place simultaneously''.
The term parallel algorithm was introduced after the first studies on
parallelization of `classical' algorithms (see e.g.~\cite{ET63a}).
One of the earliest uses of the term in the computer science literature 
was in~\cite{Rei68a}. 
In that paper, a formalization of the notion of a parallel algorithm is
given that does not depend on a particular machine model.
However, the formalization is far from covering everything that is 
currently considered a parallel algorithm.

Since the introduction of the first models of parallel computation, it 
is common practice to identify parallel algorithms with the abstract 
machines considered in a particular model of parallel computation.
Many adjustments of early models based on random access machines have 
been proposed, in particular of those introduced 
in~\cite{CZ89a,FW78a,Gib89a,Gol78a}.
The resulting wide variety of proposed models of parallel computation 
does not make it easier to come up with a formal notion of a parallel 
algorithm that encompasses everything considered a parallel algorithm.
It therefore seems useful to start with distinguishing different types 
of parallel algorithms and generalizing the notion of a proto-algorithm 
to a notion of a parallel algorithm per type of parallel algorithms.

The generalization of the notion of a proto-algorithm to a notion of an 
interactive proto-algorithm is not so easy too.
As with parallel proto-algorithms, the main reason for this is that 
there is no consensus on which properties are essential for an 
interactive algorithm.
An interactive algorithm is usually informally described by a sentence 
like ``An interactive algorithm is an algorithm that can interact with 
the environment in which it takes place''.
In~\cite{BG06a}, a specific view on the nature of interactive algorithms 
is discussed in detail, culminating in a characterization of interactive 
algorithms by a number of postulates.
This view is the only one found in the computer science literature so 
far. 
Some of its details are based on choices whose impact on the generality 
of the characterization is not clear.

Recently, several models of interactive computation have been proposed.
They are based on variants of Turing machines, to wit interactive Turing 
machines~\cite{LW01a}, persistent Turing machines~\cite{GSAS04a}, and 
reactive Turing machines~\cite{BLT13a}.
These models are closely related.
In~\cite{BLT13a}, it is established that reactive Turing machines are at 
least as expressive as persistent Turing machines.
Moreover, it is established in that paper that the behaviour of a 
reactive Turing machine can be defined by a recursive specification in a 
process algebra closely related to \ACPet~\cite[Section~5.3]{BW90}, an 
extension of \BPAde\ that includes parallel composition.

\section{Concluding Remarks}
\label{sect-conclusions}

I have reported on a quest for a satisfactory formalization of the 
notion of an algorithm.
I have introduced the notion of a proto-algorithm.
Algorithms are expected to be equivalence classes of proto-algorithms 
under an appropriate equivalence relation.
I have defined three equivalence relations on proto-algorithms.
Two of them give bounds between which an 
appropriate equivalence relation must lie.
The third one, called algorithmic equivalence, lies in between these two 
and is likely an appropriate one.
I have also presented a sound method for proving algorithmic equivalence 
of two proto-algorithms using the imperative process algebra \deBPAder.

The notion of a proto-algorithm defined in this paper does not depend on 
any particular machine model or algorithmic language, and also has most 
properties that are generally considered to belong to the important 
ones of an algorithm.
This makes it neither too concrete nor too abstract to be a appropriate 
basis for investigating what exactly an algorithm is in the setting of 
emerging types of computation, such as interactive computation.

Due to the connection between proto-algorithms and processes that is 
expressed by Theorem~\ref{theorem-grtopr-algeqv}, 
\deACPetr~\cite{Mid21a}, an extension of \deBPAder\ that includes among 
other things parallel composition, is potentially a suitable tool to 
find out how to generalize the notion of a proto-algorithm to the 
different types of parallel algorithms.

\appendix

\section*{Appendix: Axioms of \deBPAder}
\label{appendix}

The axioms of \deBPAder\ are presented in Table~\ref{axioms-deBPAder}. 
In this table, 
$t$ stands for an arbitrary term from $\ProcTerm$, 
$\phi$ and $\psi$ stand for arbitrary terms from $\CondTerm$,
$e$ stands for an arbitrary term from $\DataTerm$,
$a$ stands for an arbitrary basic action from $\Act$,
$v$ stands for an arbitrary flexible variable from $\FlexVar$, 
$\rho$ stands for an arbitrary flexible variable valuation from 
$\FVarVal$,
$X$ stands for an arbitrary variable from $\cX$, 
$S$ stands for an arbitrary linear recursive specification over 
\deBPAde.
The notation $\rec{t}{S}$ is used in axiom RDP for $t$ with, for all 
$X \in \vars(S)$, all occurrences of $X$ in $t$ replaced by 
$\rec{X}{S}$.
The homomorphic extensions of a flexible variable valuation $\rho$ from 
$\FlexVar$ to $\DataTerm$ and $\CondTerm$ are denoted in axioms V3 and 
V5 by $\rho$ as well.
The notation $\rho\mapupd{e}{v}$ is used in axiom V3 for the flexible 
variable valuation $\rho'$ defined by $\rho'(v') = \rho(v')$ if 
$v' \neq v$ and $\rho'(v) = e$.
\begin{table}[!hbt]
\caption{Axioms of \deBPAder}
\label{axioms-deBPAder}
\begin{eqntbl}
\begin{axcol}
x \altc y = y \altc x                                & & \axiom{A1} \\
(x \altc y) \altc z = x \altc (y \altc z)            & & \axiom{A2} \\
x \altc x = x                                        & & \axiom{A3} \\
(x \altc y) \seqc z = x \seqc z \altc y \seqc z      & & \axiom{A4} \\
(x \seqc y) \seqc z = x \seqc (y \seqc z)            & & \axiom{A5} \\
x \altc \dead = x                                    & & \axiom{A6} 
\\[1ex]
\True \gc x = x                                      & & \axiom{GC1} \\
\False \gc x = \dead                                 & & \axiom{GC2} \\
\phi \gc \dead = \dead                               & & \axiom{GC3} \\
\phi \gc (x \altc y) = \phi \gc x \altc \phi \gc y   & & \axiom{GC4} \\
\phi \gc x \seqc y = (\phi \gc x) \seqc y            & & \axiom{GC5} \\
\phi \gc (\psi \gc x) = (\phi \Land \psi) \gc x      & & \axiom{GC6} \\
(\phi \Lor \psi) \gc x = \phi \gc x \altc \psi \gc x & & \axiom{GC7}
\end{axcol}  
\quad\;\;\;
\begin{axcol}
\dead \seqc x = \dead                                & & \axiom{A7} \\
x \seqc \ep = x                                      & & \axiom{A8} \\
\ep \seqc x = x                                      & & \axiom{A9} \\
\\  
\rec{X}{S} = \rec{t}{S}      & \mif X \!= t \;\in\, S & \axiom{RDP} \\
S \Limpl X = \rec{X}{S}      & \mif X \in \vars(S)    & \axiom{RSP} 
\\[1ex]
\eval{\rho}(\ep) = \ep                               & & \axiom{V1} \\
\eval{\rho}(a \seqc x) = a \seqc \eval{\rho}(x)
                                                     & & \axiom{V2} \\
\multicolumn{2}{l}{
\eval{\rho}(\ass{v}{e} \seqc x) = 
{\ass{v}{\rho(e)} \seqc \eval{\rho\mapupd{\rho(e)}{v}}(x)} 
                                                     } & \axiom{V3} \\
\eval{\rho}(x \altc y) = \eval{\rho}(x) \altc \eval{\rho}(y)
                                                     & & \axiom{V4} \\
\eval{\rho}(\phi \gc y) = \rho(\phi) \gc \eval{\rho}(x)
                                                     & & \axiom{V5} \\
e = e'         & \mif \Sat{\gD}{e = e'}                & \axiom{IMP1} \\
\phi = \psi    & \mif \Sat{\gD}{\phi \Liff \psi}       & \axiom{IMP2} 
\end{axcol}
\end{eqntbl}
\end{table}

\bibliographystyle{splncs04}
\bibliography{ALG}

\end{document}